\documentclass[final,5p,times]{elsarticle}

\usepackage{lineno,hyperref}
\usepackage{comment}
\modulolinenumbers[5]

\journal{Physics Letters A}

\usepackage{amssymb}
\usepackage{amsmath}
\usepackage{amsfonts}
\usepackage{physics}

\begin{document}

\begin{frontmatter}




\title{The impact of nonlocal coupling on deterministic and stochastic wavefront propagation in an ensemble of bistable oscillators}


\author{Vladimir V. Semenov}
\ead{semenov.v.v.ssu@gmail.com}

\address{Institute of Physics, Saratov State University, 83 Astrakhanskaya Street, Saratov, 410012, Russia}

\date{\today}

\begin{abstract}
Based on methods of numerical simulation, the constructive role of nonlocal coupling is demonstrated in the context of wavefront propagation observed in an ensemble of overdamped bistable oscillators. Firstly, it is shown that the wavefront propagation can be controlled, i.e. accelerated or slowed down, by varying the strength and radius of nonlocal interactions. This applies to both deterministic and stochastic wavefront propagation. Secondly, nonlocal interactions are found to facilitate preservation of spatial domains and fronts being totally destroyed due to the action of noise in the case of local coupling. In addition, a new finding concerning the action of additive noise is reported: it is shown that additive noise is capable of accelerating front propagation if the local dynamics involves asymmetry.
\end{abstract}



\begin{keyword}
Wavefront propagation \sep nonlocal coupling \sep bistability \sep noise


\PACS 05.10.-a \sep 05.45.-a \sep 05.40.Ca

\end{keyword}

\end{frontmatter}

\section{Introduction}
\label{intro}

Wavefront propagation in bistable media can be manifested in various forms from simple ones associated with single fronts travelling in 1D-space \cite{rinzel1982,engel1985,garcia-ojalvo1999,mendez2011} to more complicated realizations in multi-dimensional space corresponding to phase separation accompanied by further domain growth and coarsening \cite{cugliandolo2010,giacomelli2012}. Such effects are a frequent occurrence in chemistry \cite{schloegl1972,schloegl1983,loeber2014}, flame propagation theory \cite{zeldovich1938}, electronics \cite{loecher1998,semenov2018,zakharova2024}, optics \cite{giacomelli2012,semenov2023}, population statistics \cite{mendez2011}, physics of magnetic phenomena \cite{cugliandolo2010,caccioli2008}. Models of wavefront propagation involving local bistability are not confined only to reaction-diffusion systems, but also include populations and networks of coupled oscillators \cite{zakharova2023,semenov2023-2} and delayed-feedback oscillators \cite{giacomelli2012,semenov2018,zakharova2024}. 

It is well-known that the presence of asymmetry in bistable spatially-extended systems can be a reason for the wavefront propagation. Moreover, the bigger is the asymmetry, the faster are the propagating fronts. Thus, symmetry control schemes can be applied for controlling the front propagation velocity and direction \cite{semenov2023}. In the presence of noise, the wavefront propagation control can be realized by varying the amplitude of the stochastic impact \cite{schimansky1983,engel1985,loecher1998}. In particular, multiplicative noise is often used to realize the impact on the systematic part of the front dynamics \cite{engel1985,garcia-ojalvo1999,mendez2011}. 

Multiplicative noise is well-known in the context of the stochastic control of the wavefront propagation speed in bistable systems. Moreover, the presence of multiplicative noise can give rise to noise-induced fronts and noise-triggered wavefront propagation in such systems \cite{santos1999}. Additive noise is known to control propagating fronts in excitable media and lattices (for instance, see resonant-like regularization of noise-induced fronts in a semiconductor lattice \cite{hizanidis2006}), as well as in systems exhibiting self-sustained spatially-periodic states \cite{clerc2005}. However, additive noise is less studied as a control factor for propagating fronts associated with the property of bistability. In particular, it has been reported in Ref. \cite{engel1985} that additive noise does not induce wavefront propagation in a symmetric bistable medium. In contrast, it is demonstrated in the present paper that additive noise can speed up the fronts in a bistable model naturally propagating in the absence of noise due to the presence of asymmetry.

Adjusting the coupling strength and modifying the coupling topology represents a distinguishable approach for the wavefront propagation control which can be implemented in ensembles and networks of coupled oscillators. In particular, the property of the wavefront propagation in multiplex networks was reported in recent paper \cite{semenov2023-2}. In addition, the properties of nonlocal coupling were applied to control the activity bump expansion in an ensemble of bistable spiking oscillators \cite{semenov2023}. The effects revealed in Ref. \cite{semenov2023} indicate the constructive role of nonlocal coupling in the context of the wavefront propagation resulting in expansion of spiking activity patterns. Moreover, it was demonstrated that introducing nonlocal couplings can prevent wavefront patterns from being destroyed by noise.
However, the ensemble models considered in paper \cite{semenov2023} are characterised by the complex dynamics of the ensemble element. It is not clear whether the established phenomena have general character or are caused by the intrinsic peculiarities of the local dynamics. To answer this question, the wavefront propagation is discussed in the current paper on an example of nonlocally coupled oscillators exhibiting the simplest kind of bistability: the coexistence of two steady states.

A distinguishable aim of the present work is to extend a manifold of phenomena associated with the presence of nonlocal character of spatial interaction \cite{andreu2010,nicola2002,sheintuch1997,bordyugov2006,colet2014,gelens2014}. For instance, nonlocal coupling is known to induce a wide spectrum of spatiotemporal effects observed in chemical systems \cite{mazouz1997} including various manifestations of the Belousov-Zhabotinsky reaction \cite{hildebrand2001,nicola2006}, magnetic fluids \cite{friedrichs2003}, and is responsible for the exhibition of fingering \cite{pismen2017}, remote wave triggering \cite{christoph1999}, Turing structures \cite{li2001}, periodic travelling waves, single and multiple pulses \cite{alfaro2014,volpert2015,achleitner2015}, wave instabilities \cite{nicola2006}, or spatiotemporal chaos \cite{varela2005}. In the context of ensembles of coupled oscillators, the nonlocal coupling is known to provide for observation of patched patterns \cite{franovic2022-2},
chimera \cite{abrams2004,zakharova2020,parastesh2021,franovic2021} and solitary \cite{jaros2018,berner2020,schuellen2022,franovic2022} states. The nonlocal diffusion is successfully used for the wave propagation control in excitable \cite{bachmair2014} and bistable \cite{colet2014,gelens2014,siebert2014,siebert2015} media. Complementing these results, the current paper addresses waves in ensembles of bistable systems with nonlocal interactions focusing on controlling wavefront velocity in the noiseless and stochastic cases and saving propagating fronts destroyed by noise in the presence of local coupling. 

\section{Model and methods}
\label{model_and_methods}
The model under study represents an ensemble of nonlocally coupled identical overdamped bistable oscillators. The system equations take the following form:
\begin{equation}
\label{eq:system}
\begin{array}{l}
\dfrac{dx_{i}}{dt}=-x_{i}(x_{i}-a)(x_{i}+b+\sqrt{2D_b}n_i(t))\\
+\sqrt{2D}n_i(t) + \dfrac{\sigma}{2R}\sum\limits^{i+R}_{j=i-R}(x_{j}-x_{i}), 
\end{array}
\end{equation}
where $x_i$ are the dynamical variables, $i=1, 2, ..., N$, where $N=10^3$ is the total number of oscillators. The coupling parameters are coupling strength $\sigma$ and coupling radius $R$. Parameters $a,b>0$ define whether the nonlinearity of the ensemble elements is symmetric ($a=b$) or asymmetric ($a\neq b$). Model (\ref{eq:system}) is studied by means of numerical simulations. In more detail, the differential equations are integrated using the Heun method \cite{mannella2002} with the time step $\Delta t=0.001$. The initial conditions are chosen to induce two wavefronts: $x_i(t_0=0)=-b$ for $i \in [250:750]$ and $x_i(t_0=0)=a$ otherwise. The boundary conditions are periodic, i.e. indices 
$j<1$ and $j>N$ are changed to $j+N$ and $j-N$ correspondingly. If the ensemble elements are symmetric, the obtained fronts do not move and their position is fully determined by the initial conditions. In contrast, the fronts can propagate over the ensemble in a case of asymmetric oscillators. 

Studying deterministic ensemble is complemented by the exploration of the stochastic system. Two types of noise are considered: multiplicative ($D=0$, $D_b\neq 0$) or additive ($D\neq0$, $D_b=0$). In both cases, each oscillator $x_i$ contains a statistically independent source of white Gaussian noise of intensity $D$, i.e., $<n_i(t)>=0$ and $<n_i(t)n_{j}(t)>=\delta_{ij}\delta(t-t')$, $\forall i,j$.

To visualise the system's evolution, the space-time plots $x_i(t)$ are used. In addition, the front propagation velocity is introduced as $v=(w(t) - w(t+\Delta t))/2\Delta t$. Here, $w(t)$ and $w(t+\Delta t)$ are the widths (integer numbers being a result of the index subtraction) of the central spatial domain corresponding to the state $x_i(t)=-b$ at the moments $t$ and  $t+\Delta t$, respectively (illustrated in Fig.~\ref{fig1}~(c)). Note that the velocity $v$ actually applies to two fronts propagating in the opposite directions, cf. arrows in Fig.~\ref{fig1}~(a)). The related formula uses two fronts to reduce inaccuracies, and factor 2 is introduced to single out one of the fronts. In summary, quantity $v$ is a velocity of the left front in Fig.~\ref{fig1}~(a) propagating to the right and, similarly, the velocity of the right front moving to the left. 

The total integration time used to build the space-time plots and to calculate the wavefront propagation velocity was $t_{\text{total}}=10^4$. In the presence of noise, the wavefront propagation is characterised by fluctuations of the wavefront instantaneous position, especially for noise of large intensity. For this reason, the mean wavefront propagation velocity was used to analyze the stochastic ensemble dynamics. The mean wavefront propagation velocity $<v>$ was obtained as an average of 40 values of $v$ calculated on the base of 40 numerical simulations started from the same initial conditions.

\section{Deterministic effects}
\label{sec_deterministic}

\begin{figure}[t]
\centering
\includegraphics[width=0.48\textwidth]{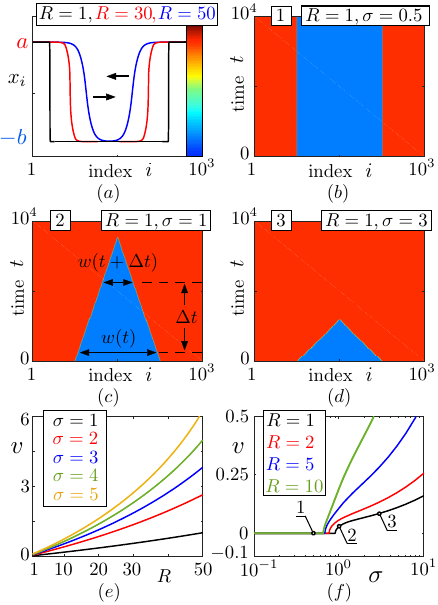}
\caption{Control of deterministic wavefront propagation when varying the coupling strength and radius (see Eqs.~(\ref{eq:system}) at $D=D_b=0$): (a) Evolution of the wave profile when increasing the coupling radius at $\sigma=1$; (b)-(d) Space-time plots illustrating the front propagation at $R=1$ and increasing coupling strength; (e) Dependencies of the wavefront propagation velocity on the coupling radius obtained for different coupling strength; (f) Dependencies of the wavefront propagation velocity on the coupling strength obtained at different coupling radius. Points 1-3 in panel (f) correspond to space-time plots in panels (b)-(d). Other parameters are $a=1$, $b=0.9$.}
\label{fig1}
\end{figure}
First, the impact of nonlocal coupling is discussed on an example of an ensemble of asymmetric deterministic oscillators: $a=1$, $b=0.9$, $D_b=D=0$. For chosen initial conditions (see the previous section), ensemble (\ref{eq:system}) is prepared in an inhomogeneous state including both steady states such that two spatial domains $x_i(t)=a$ and $x_i(t)=-b$ are formed [Fig. ~\ref{fig1}~(a)]. In the presence of weak coupling, the fronts separating the domains do not move [Fig. ~\ref{fig1}~(b)]. When the coupling strength increases and passes through certain critical value, the fronts start to propagate such that spatial state $x_i(t)=a$ begins to dominate and invades the entire space [Fig. ~\ref{fig1}~(c)]. This process gets accelerated as $\sigma$ is increased [Fig. \ref{fig1}~(d)]. 

Increasing the coupling radius results in smoothing the shape of propagating fronts (the wavefronts obtained at different $R$ are depicted Fig.~\ref{fig1}~(a)) and increasing the front propagation velocity calculated at fixed coupling strength [Fig.~\ref{fig1}~(e)]. This fact is also illustrated in Fig.~\ref{fig1}~(f) as the evolution of the dependence of the wavefront propagation velocity on the coupling strength when increasing the coupling radius. It is important to note that tuning the coupling radius allows to shift the threshold values of the coupling strength corresponding to starting the wavefront propagation (see Fig.~\ref{fig1}~(f)). 


\section{Stochastic dynamics}
\subsection{The impact of multiplicative noise}
\label{sec_multiplicative}
\begin{figure}[t]
\centering
\includegraphics[width=0.48\textwidth]{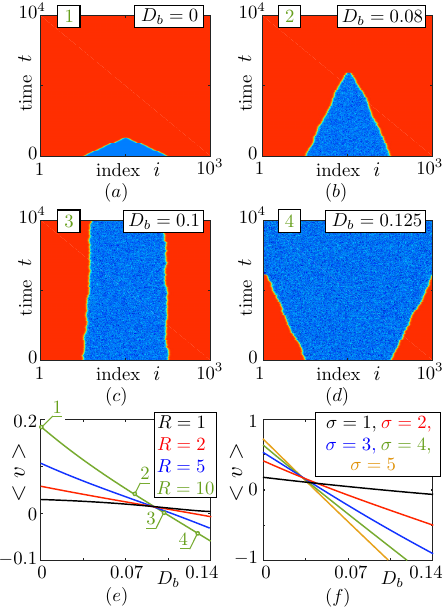}
\caption{Control of stochastic wavefront propagation in ensemble (\ref{eq:system}) ($D=0$, $D_b \neq 0$) by varying the coupling parameters: (a)-(d) Space-time plots obtained at $\sigma=1$, $R=10$ and varying noise intensity; (e) Dependencies of the mean wavefront propagation velocity on the noise intensity obtained at fixed coupling strength, $\sigma=1$, and different coupling radius.
Points 1-4 in panel (e) correspond to the space-time plots in panels (a)-(d); (f) Dependencies of the mean wavefront propagation velocity on the noise intensity obtained at fixed coupling radius, $R=10$, and varying coupling strength. Other parameters are $a=1$, $b=0.9$.}
\label{fig2}
\end{figure}

As already mentioned in the introduction, wavefront propagation can be controlled by applying the multiplicative noise, i.e. the front propagation velocity depends on the noise intensity. This effect also takes place in the presence of nonlocal coupling. As illustrated in Fig.~\ref{fig2}~(a)-(d) on an example of nonlocally coupled oscillators in the presence of parametric noise (see Eqs. \ref{eq:system} at $D_b\neq0$, $D=0$), the wavefront propagation slows down with the noise intensity growth, so that the expansion of the state $x_i(t)=a$ stops, see Fig.~\ref{fig2}~(c). Further increasing the noise strength inverts the front propagation and the quiescent steady state regime $x_i(t)=-b$ invades the whole space [Fig.~\ref{fig2}~(d)]. Thus, the multiplicative noise-based control of the wavefront propagation in the presence of local and nonlocal coupling does not qualitatively differ from each other. However, the quantitative difference takes place. As demonstrated in Fig.~\ref{fig2}~(e), increasing the coupling radius at fixed noise intensity and coupling strength leads to faster propagation, similarly to the effects observed in the deterministic system. In addition, the dependencies $<v(D_b)>$ obtained at fixed coupling radius and varying coupling strength [Fig.~\ref{fig2}~(f)] clearly indicate that growth of the coupling strength makes the stochastic wavefront propagation faster, similarly to the impact on the deterministic process. 

\subsection{Role of additive noise}
\label{sec_additive}

\begin{figure}[t]
\centering
\includegraphics[width=0.48\textwidth]{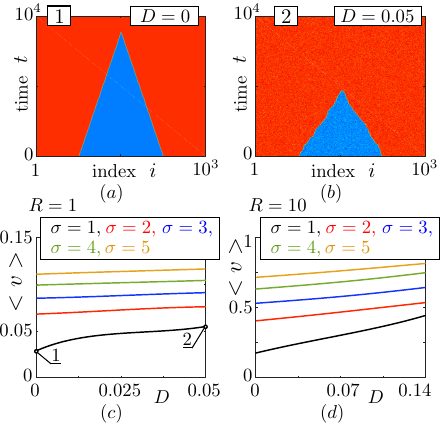}
\caption{Additive noise accelerates wavefront propagation in the presence of asymmetry: panels (a) and (b) are space-time plots obtained at $D=0$ and $D=0.05$ and fixed coupling parameters $\sigma=1$, $R=1$. Panels (c) and (d) illustrate the dependence of the mean propagation velocity $<v>$ on the noise intensity $D$ when increasing the coupling strength $\sigma$ in the presence of local ($R=1$, see panel (c)) and nonlocal ($R=10$, see panel (d)) interaction. Other parameters are $a=1$, $b=0.9$, $D_b=0$.}
\label{fig3}
\end{figure}

As mentioned in Ref. \cite{engel1985}, additive noise does not induce the wavefront propagation in symmetric bistable reaction-diffusion systems. This rule is also fulfilled in ensemble (\ref{eq:system}) of coupled bistable oscillators with symmetric nonlinearity in the absence of multiplicative noise ($a=b$, $D_b=0$) similarly to paper \cite{engel1985}: the mean wavefront propagation velocity remains to be zero when increasing intensity $D$. In contrast, additive noise can impact the propagating fronts when the interacting oscillators are asymmetric. In such a case, additive noise speeds up the fronts as demonstrated in Fig.~\ref{fig3}~(a),(b). This effect is observed in the presence of local coupling ($R=1$, see Fig.~\ref{fig3}~(c)) as well as when increasing the coupling radius [Fig.~\ref{fig3}~(d)]. When the coupling strength increases, the influence of additive noise becomes less visible, but persists. Similarly to the  ensemble in the presence of multiplicative noise and to the deterministic one, 
 propagation velocity grows with the coupling radius [Fig.~\ref{fig3}~(c),(d)]. 

\begin{figure}[t]
\centering
\includegraphics[width=0.48\textwidth]{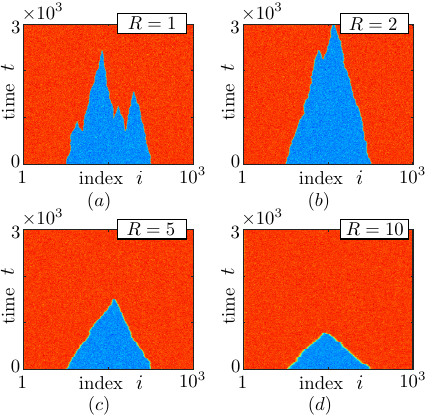}
\caption{Increasing the coupling radius prevents wavefront propagation failure induced by additive noise ($D=0.07$, $D_b=0$ in Eqs. (\ref{eq:system})): the spatial domain corresponding to $x_i=-b$ is destroyed in case of local coupling (panel (a)) but persists when the interaction is nonlocal (panels (b)-(d)). Other parameters are $a=1$, $b=0.9$, $\sigma=1$.}
\label{fig4}
\end{figure}

Dependencies $<v(D)>$ in Fig. \ref{fig3}(c),(d) are presented for different ranges of $D$. This is due to the fact that noise can destroy the propagating fronts in the presence of local coupling. 
Such destruction of propagating fronts induced by additive noise is observed at lower noise intensities as compared to the multiplicative noise. In particular, this effect is occasionally observed in model (\ref{eq:system}) at $D_b=0$, $D>0.05$ and inevitably occurs when the additive noise intensity exceeds the value $D_{\text{critical}}\approx 0.064$ [Fig.~\ref{fig4}~(a)].  

As depicted in Fig.~\ref{fig4}~(a), stochastic force destroys the initial dynamics such that new domains and fronts eventually appear and disappear. However, the propagating fronts obtained from the same initial conditions and at the same noise intensity sustain when increasing the coupling radius [Fig.~\ref{fig4}~(b)]. Similarly to the previous examples of the wavefront propagation illustrated in Figs.~\ref{fig1} - \ref{fig3}, growth of the coupling radius gives rise to larger wavefront velocities and the exhibited front propagation finishes earlier [Fig.~\ref{fig4}~(c),(d)].

\section{Conclusion}
\label{sec:conclusion}
A constructive role of nonlocal coupling has been demonstrated in the current paper in the context of the deterministic and stochastic wavefront propagation control realized in a simple model of interacting bistable oscillators. In particular, it has been established that increasing the coupling radius at fixed coupling strength speeds up the propagating fronts which was observed both in the presence and in the absence of noise. This result is in a good correspondence with Ref. \cite{siebert2014} where the action of nonlocal spatial interaction on the wavefront propagation in a deterministic bistable medium was explained by means of analytical solution. A comparative analysis of the present results and materials published in Ref. \cite{zakharova2023}, where the similar effects are reported, suggests that such impact of nonlocal coupling has a general character and is independent on the dynamics of single elements.

The second important impact of nonlocality of interactions consists in preventing wavefront propagation failure in the presence of noise, which inevitably occurs if the interactions are local. The same effect is emphasized in paper \cite{zakharova2023} on example of more complicated phenomena related to the wavefront propagation. This indicates the fact that the properties of nonlocal interaction can be applied to save fronts in a wide spectrum of ensembles and media of different nature. 

It is important to note the distinctive peculiarities of the wavefront propagation revealed in the system under study. In contrast to bistable media, an ensemble of deterministic bistable oscillators is a model with discretized space and exhibits the wavefront propagation after the coupling strength exceeds certain critical value despite the presence of asymmetry. The second fact refers to the stochastic wavefront propagation: the dependencies of the mean wavefront velocity on the noise intensity illustrated in the present paper qualitatively differ from ones discussed in \cite{zakharova2023}. Comparison of the current results with the materials from Ref. \cite{zakharova2023} suggests that the impact of noise on the mean propagation velocity (monotonic or non-monotonic, linear or nonlinear, the presence of the flat areas, etc.) depends on the character of local dynamics. In particular, elements of the ensemble considered in Ref. \cite{zakharova2023} exhibit the coexistence of a stable steady state and a stable limit cycle which is principally different from the bistability of the coupled elements in model (\ref{eq:system}).

Finally, it has been found that additive noise can accelerate front propagation for asymmetric local dynamics. This result distinguishes the present paper from most of publications addressing the problem of stochastic wavefront propagation and focused on the action of multiplicative noise (for instance, see Refs. \cite{engel1985,garcia-ojalvo1999,mendez2011}). This intriguing issue could be a subject of future study as well as studying the impact of another kinds of topology, for instance, higher-order interactions usually considered in the context of synchronization \cite{dai2020,kovalenko2021,boccaletti2023}.

The obtained results can be useful in the context of many applications involving control of propagating fronts and spatial domains from spread of infections and diseases to signal transmission and control of animal population dynamics. A distinguishable application where the property of bistability and the symmetry control are of particular importance is physical implementation of spin-networks (or spin-glasses) as solvers of combinatorial optimization problems \cite{boehm2019,boehm2021,mohseni2022}. The implemented spins are assumed to exhibit two symmetric, opposite states (spin-up and spin-down). In such a case, non-zero wavefront propagation velocity observed in the network prepared in an inhomogeneous state including both steady states can indicate the presence of asymmetry. Indeed, static wavefronts (or almost static) are usually observed in symmetric bistable systems (or almost symmetric). Thus, one can expect that the case of minimal wavefront propagation speed corresponds to the best performance of the spin-network. If certain physical implementation of spin-networks principally involves the presence of nonlocal interactions, this could be used as an additional factor for the symmetry control.

\section*{Declaration of Competing Interest}
The author declares that he has no known competing financial interests or personal relationships that could have appeared to influence the work reported in this paper. 

\section*{Acknowledgments}
This work was supported by the Russian Science Foundation (project No. 24-72-00054).


\begin{thebibliography}{10}
\expandafter\ifx\csname url\endcsname\relax
  \def\url#1{\texttt{#1}}\fi
\expandafter\ifx\csname urlprefix\endcsname\relax\def\urlprefix{URL }\fi
\expandafter\ifx\csname href\endcsname\relax
  \def\href#1#2{#2} \def\path#1{#1}\fi

\bibitem{rinzel1982}
J.~Rinzel, D.~Terman, Propagation phenomena in a bistable reaction-diffusion
  system, SIAM Journal on Applied Mathematics 42~(5) (1982) 1111--1137.

\bibitem{engel1985}
A.~Engel, Noise-induced front propagation in a bistable system, Phys. Lett. A
  113~(3) (1985) 139--142.

\bibitem{garcia-ojalvo1999}
J.~Garcia-Ojalvo, J.~Sancho, Noise in Spatially Extended Systems, Springer,
  1999.

\bibitem{mendez2011}
V.~M{\'e}ndez, I.~Llopis, D.~Campos, W.~Horsthemke, Effect of environmental
  fluctuations on invasion fronts, J. Theor. Biol. 281~(1) (2011) 31--38.

\bibitem{cugliandolo2010}
L.~Cugliandolo, Topics in coarsening phenomena, Physica A 389~(20) (2010)
  4360--4373.

\bibitem{giacomelli2012}
G.~Giacomelli, F.~Marino, M.~Zaks, S.~Yanchuk, Coarsening in a bistable system
  with long-delayed feedback, Europhys. Lett. 99~(5) (2012) 58005.

\bibitem{schloegl1972}
F.~Schl{\"o}gl, Chemical reaction models for non-equilibrium phase transitions,
  Zeitschrift f{\"u}r Physik 253~(2) (1972) 147--161.

\bibitem{schloegl1983}
F.~Schl{\"o}gl, C.~Escher, R.~Berry, Fluctuations in the interface between two
  phases, Phys. Rev. A 27~(5) (1983) 2698--2704.

\bibitem{loeber2014}
A.~Mikhailov, G.~Ertl (Eds.), Engineering of Chemical Complexity II, World
  Scientific Lecture Notes in Complex Systems, World Scientific, 2014, Ch.
  Control of Chemical Wave Propagation, pp. 185--207.

\bibitem{zeldovich1938}
Y.~Zel'dovich, D.~Frank-Kamenetskii, Theory of uniform flame propagation, Dokl.
  Akad. Nauk SSSR 19 (1938) 693--798.

\bibitem{loecher1998}
M.~L{\"o}cher, D.~Cigna, E.~Hunt, Noise sustained propagation of a signal in
  coupled bistable electronic elements, Phys. Rev. Lett. 80~(23) (1998)
  5212--5215.

\bibitem{semenov2018}
V.~Semenov, Y.~Maistrenko, Dissipative solitons for bistable delayed-feedback
  systems, Chaos 28~(10) (2018) 101103.

\bibitem{zakharova2024}
A.~Zakharova, V.~Semenov, Delayed-feedback oscillators replicate the dynamics
  of multiplex networks: wavefront propagation and stochastic resonance, Neural
  Networks 183 (2025) 106939.

\bibitem{semenov2023}
V.~Semenov, X.~Porte, L.~Larger, D.~Brunner, Deterministic and stochastic
  coarsening control in optically-addressed spatial light modulators subject to
  optical feedback, Phys. Rev. B 108~(2) (2023) 024307.

\bibitem{caccioli2008}
F.~Caccioli, S.~Franz, M.~Marsili, Ising model with memory: coarsening and
  persistence properties, Journal of Statistical Mechanics: Theory and
  Experiment 2008~(7) (2008) P07006.

\bibitem{zakharova2023}
A.~Zakharova, V.~Semenov, Stochastic control of spiking activity bump
  expansion: monotonic and resonant phenomena, Chaos 33~(8) (2023) 081101.

\bibitem{semenov2023-2}
V.~Semenov, S.~Jalan, A.~Zakharova, Multiplexing-based control of wavefront
  propagation: the interplay of inter-layer coupling, asymmetry and noise,
  Chaos, Solitons and Fractals 173 (2023) 113656.

\bibitem{schimansky1983}
L.~Schimansky-Geyer, A.~Mikhailov, W.~Ebeling, Effect of fluctuation on plane
  front propagation in bistable nonequilibrium systems, Annalen der Physik
  495~(4-5) (1983) 277--286.

\bibitem{santos1999}
M.~Santos, J.~Sancho, Noise-induced fronts, Phys. Rev. E 59~(1) (1999) 98--102.

\bibitem{hizanidis2006}
J.~Hizanidis, A.~Balanov, A.~Amann, E.~Sch{\"o}ll, Noise-induced front motion:
  Signature of a global bifurcation, Phys. Rev. Lett. 96~(24) (2006) 244104.

\bibitem{clerc2005}
M.~Clerc, C.~Falcon, E.~Tirapegui, Additive noise induces front propagation,
  Phys. Rev. Lett. 94~(14) (2005) 148302.

\bibitem{andreu2010}
F.~Andreu-Vaillo, J.~Maz{\'o}n, J.~Rossi, J.~Toledo-Melero, Nonlocal Diffusion
  Problems, American Mathematical Society, 2010.

\bibitem{nicola2002}
E.~Nicola, M.~Or-Guil, W.~Wolf, M.~B{\"a}r, Drifting pattern domains in a
  reaction-diffusion system with nonlocal coupling, Phys. Rev. E 65~(5) (2002)
  055101(R).

\bibitem{sheintuch1997}
M.~Sheintuch, O.~Nekhamkina, Reaction-diffusion patterns on a disk or a square
  in a model with long-range interaction, J. Chem. Phys. 107~(19) (1997)
  8165--8174.

\bibitem{bordyugov2006}
G.~Bordyugov, H.~Engel, Creating bound states in excitable media by means of
  nonlocal coupling, Phys. Rev. E 74~(1) (2006) 016205.

\bibitem{colet2014}
P.~Colet, M.~Mat{\'\i}as, L.~Gelens, D.~Gomila, Formation of localized
  structures in bistable systems through nonlocal spatial coupling. I. General
  framework, Phys. Rev. E 89~(1) (2014) 012914.

\bibitem{gelens2014}
L.~Gelens, M.~Mat{\'\i}as, D.~Gomila, T.~Dorissen, P.~Colet, Formation of
  localized structures in bistable systems through nonlocal spatial coupling.
  II. The nonlocal ginzburg-landau equation, Phys. Rev. E 89~(1) (2014) 012915.

\bibitem{mazouz1997}
N.~Mazouz, G.~Fl{\"a}tgen, K.~Krisher, Tuning the range of spatial coupling in
  electrochemical systems: From local via nonlocal to global coupling, Phys.
  Rev. E 55~(3) (1997) 1997.

\bibitem{hildebrand2001}
M.~Hildebrand, H.~Sk{\o}dt, K.~Sowalter, Spatial symmetry breaking in the
  Belousov-Zhabotinsky reaction with light-induced remote communication, Phys.
  Rev. Lett. 87~(8) (2001) 088303.

\bibitem{nicola2006}
E.~Nicola, M.~B{\"a}r, H.~Engel, Wave instability induced by nonlocal spatial
  coupling in a model of the light-sensitive Belousov-Zhabotinsky reaction,
  Phys. Rev. E 73~(6) (2006) 2006.

\bibitem{friedrichs2003}
R.~Friedrichs, A.~Engel, Non-linear analysis of the Rosensweig instability,
  Europhys. Lett. 63~(6) (2003) 826--832.

\bibitem{pismen2017}
L.~Pismen, Encyclopedia of Complexity and Systems Science, Springer, 2017, Ch.
  Patterns and Interfaces in Dissipative Dynamics, pp. 1--21.

\bibitem{christoph1999}
J.~Christops, R.~Otterstedt, M.~Eiswirth, N.~Jaeger, J.~Hudson, Negative
  coupling during oscillatory pattern formation on a ring electrode, J. Chem.
  Phys. 110~(17) (1999) 8614--8621.

\bibitem{li2001}
Y.~Li, J.~Oslonovitch, N.~Mazouz, F.~Plence, K.~Krisher, G.~Ertl, Turing-type
  patterns on electrode surfaces, Science 291~(5512) (2001) 2395--2398.

\bibitem{alfaro2014}
M.~Alfaro, J.~Coville, G.~Raoul, Bistable travelling waves for nonlocal
  reaction diffusion equations, Discrete and Continuous Dynamical Systems
  34~(5) (2014) 1775--1791.

\bibitem{volpert2015}
V.~Volpert, Pulses and waves for a bistable nonlocal reaction--diffusion
  equation, Applied Mathematics Letters 44 (2015) 21--25.

\bibitem{achleitner2015}
F.~Achleitner, C.~Kuehn, Traveling waves for a bistable equation with nonlocal
  diffusion, Adv. Differential Equations 20~(9/10) (2015) 887--936.

\bibitem{varela2005}
H.~Varela, C.~Beta, A.~Bonnefont, K.~Krisher, Transitions to electrochemical
  turbulence, Phys. Rev. Lett. 94~(17) (2005) 174104.

\bibitem{franovic2022-2}
I.~Franovi{\'c}, S.~Eydam, Patched patterns and emergence of chaotic interfaces
  in arrays of nonlocally coupled excitable systems, Chaos 32~(9) (2022)
  091102.

\bibitem{abrams2004}
D.~Abrams, S.~Strogatz, Chimera states for coupled oscillators, Phys. Rev.
  Lett. 93~(17) (2004) 174102.

\bibitem{zakharova2020}
A.~Zakharova, Chimera patterns in networks, Springer, 2020.

\bibitem{parastesh2021}
F.~Parastesh, S.~Jafari, H.~Azarnoush, Z.~Shahriari, Z.~Wang, S.~Boccaletti,
  M.~Perc, Chimeras, Physics Reports 898 (2021) 1--114.

\bibitem{franovic2021}
I.~Franovi{\'c}, O.~Omel'chenko, M.~Wolfrum, Bumps, chimera states, and Turing
  patterns in systems of coupled active rotators, Phys. Rev. E 104~(5) (2021)
  L052201.

\bibitem{jaros2018}
P.~Jaros, S.~Brezetsky, R.~Levchenko, D.~Dudkowski, T.~Kapitaniak,
  Y.~Maistrenko, Solitary states for coupled oscillators with inertia, Chaos
  28~(1) (2018) 011103.

\bibitem{berner2020}
R.~Berner, A.~Polanska, E.~Sch{\"o}ll, S.~Yanchuk, Solitary states in adaptive
  nonlocal oscillator networks, European Physical Journal Special Topics
  229~(12) (2020) 2183--2203.

\bibitem{schuellen2022}
L.~Sch{\"u}llen, M.~Mikhailenko, E.~Medeiros, A.~Zakharova, Solitary states in
  complex networks: impact of topology, European Physical Journal Special
  Topics 231~(22) (2022) 4123--4130.

\bibitem{franovic2022}
I.~Franovi{\'c}, S.~Eydam, N.~Semenova, A.~Zakharova, Unbalanced clustering and
  solitary states in coupled excitable systems, Chaos 32~(1) (2022) 011104.

\bibitem{bachmair2014}
C.~Bachmair, E.~Sch{\"o}ll, Nonlocal control of pulse propagation in excitable
  media, European Physical Journal B 87~(11) (2014) 276.

\bibitem{siebert2014}
J.~Siebert, S.~Alonso, M.~B{\"a}r, E.~Sch{\"o}ll, Dynamics of
  reaction-diffusion patterns controlled by asymmetric nonlocal coupling as a
  limiting case of differential advection, Phys. Rev. E 89~(5) (2014) 052909.

\bibitem{siebert2015}
J.~Siebert, E.~Sch{\"o}ll, Front and Turing patterns induced by
  mexican-hat--like nonlocal feedback, Europhys. Lett. 109~(4) (2015) 40014.

\bibitem{mannella2002}
R.~Mannella, Integration of stochastic differential equations on a computer,
  International Journal of Modern Physics C 13~(9) (2002) 1177--1194.

\bibitem{dai2020}
X.~Dai, Discontinuous transitions and rhythmic states in the d-dimensional
  Kuramoto model induced by a positive feedback with the global order
  parameter, Phys. Rev. Lett. 125~(19) (2020) 194101.

\bibitem{kovalenko2021}
K.~Kovalenko, X.~Dai, K.~Alfaro-Bittner, A.~Raigorodskii, M.~Perc,
  S.~Boccaletti, Contrarians synchronize beyond the limit of pairwise
  interactions, Phys. Rev. Lett. 127~(25) (2021) 258301.

\bibitem{boccaletti2023}
S.~Boccaletti, P.~De~Lellis, C.~del Genio, K.~Alfaro-Bittner, R.~Criado,
  S.~Jalan, M.~Romance, The structure and dynamics of networks with higher
  order interactions, Physics Reports 1018 (2023) 1--64.

\bibitem{boehm2019}
F.~B{\"o}hm, G.~Verschaffelt, G.~Van~der Sande, A poor man's coherent Ising
  machine based on opto-electronic feedback systems for solving optimization
  problems, Nature Communications 10~(1) (2019) 3538.

\bibitem{boehm2021}
F.~B{\"o}hm, T.~Van~Vaerenbergh, G.~Verschaffelt, G.~Van~der Sande,
  Order-of-magnitude differences in computational performance of analog Ising
  machines induced by the choice of nonlinearity, Communications Physics 4~(1)
  (2021) 149.

\bibitem{mohseni2022}
N.~Mohseni, P.~McMahon, T.~Byrnes, Ising machines as hardware solvers of
  combinatorial optimization problems, Nature Reviews Physics 4~(6) (2022)
  363--379.

\end{thebibliography}

\end{document}